\PassOptionsToPackage{pdfpagelabels=false}{hyperref} 
\documentclass{aa}

\usepackage[varg]{txfonts}
\usepackage{graphicx}
\usepackage{natbib}
\usepackage{amssymb}
\usepackage{setspace}
\usepackage{amsmath}
\usepackage{epsfig,color}
\usepackage{multicol}
\usepackage[outdir=./]{epstopdf}

\long\def\symbolfootnote[#1]#2{\begingroup
\def\thefootnote{\fnsymbol{footnote}}\footnote[#1]{##https://www.overleaf.com/project/5b98c85301ba633150bd62a42}\endgroup} 

\title{Statistical mass function of prestellar cores from the density distribution of their natal clouds}
\author{S. Donkov$^1$ \and T. V. Veltchev$^{2, 3}$ \and
Ph. Girichidis$^4$ \and R. S. Klessen$^4$}

\institute{Department of Applied Physics, Faculty of Applied Mathematics, Technical University, 8 Kliment Ohridski Blvd., 1000 Sofia, Bulgaria; \email{savadd@tu-sofia.bg}\label{inst1} \and University of Sofia, Faculty of Physics, 5 James Bourchier Blvd., 1164 Sofia, Bulgaria; \email{eirene@phys.uni-sofia.bg}\label{inst2} \and Universit\"at Heidelberg, Zentrum f\"ur Astronomie, Institut f\"ur Theoretische Astrophysik, Albert-Ueberle-Str. 2, 69120 Heidelberg, Germany\label{inst3}
\and
Leibniz-Institut f\"ur Astrophysik Potsdam (AIP), An der Sternwarte 16, 14482 Potsdam, Germany\label{inst4}
}

\date{Received date /  Accepted date }

\begin{document}

% \label{firstpage}

\abstract{ The mass function of clumps observed in molecular clouds raises interesting theoretical issues, especially in its relation to the stellar initial mass function. We propose a statistical model of the mass function of prestellar cores (CMF), formed in self-gravitating isothermal clouds at a given stage of their evolution. The latter is characterized by the mass-density probability distribution function ($\rho$-PDF), which is a power-law with slope $q$. The variety of MCs is divided in ensembles according to the PDF slope and each ensemble is represented by a single spherical cloud. The cores are considered as elements of self-similar structure typical for fractal clouds and are modeled by spherical objects populating each cloud shell. Our model assumes relations between size, mass and density of the statistical cores. Out of them a core mass-density relationship $\rho\propto m^x$ is derived where $x=1/(1+q)$. We found that $q$ determines the existence or non-existence of a threshold density for core collapse. The derived general CMF is a power law of slope $-1$ while the CMF of gravitationally unstable cores has a slope $(-1 + x/2)$, comparable with the slopes of the high-mass part of the stellar initial mass function and of observational CMFs. }

  % scaling laws ?

\maketitle
\titlerunning{Mass function of prestellar cores}
\authorrunning{Donkov et al.}

\section{Introduction}   \label{Sec_Introduction}

Star formation is a complex, multi-scale process in the interstellar medium. Its final stages occur in the densest cloudy regions, consisting mostly of molecular gas with densities $n\gtrsim 10^{2}~{\rm cm}^{-3}$ \citep[][and the references therein]{Klessen_Glover_16}. This gas is as cold as $T\sim 10-30$~K and its thermodynamical state could be considered as approximately isothermal. An interplay between gravity, turbulence, thermal pressure and magnetic fields takes place at various scales (from thousands au to tens of pc) within these cold zones in star-forming regions. The non-thermal motions are mostly supersonic and are considered as a signature of hierarchical and chaotic collapse at all scales \citep{VS_ea_07, BP_ea_11, BP_ea_18}, accretion-driven turbulence \citep{KH_10} and/or momentum/energy deposition into the clouds by supernovae explosions \citep{Dib_Bell_Burkert_06, Padoan_ea_16} and other mechanisms. Although this physical picture is very complex, the gas dynamics at the advanced evolutionary stages of the cloud is dominated by gravity. However, the onset of a multi-scale collapse is not only determined in the densest regions. The process rather starts at Galactic scales \citep{Ibanez-Mejia_ea_16, Elmegreen_18} and continues to cascade down to smaller scales and denser regions. When gravity becomes the dominant acting force, a power-law tail (PLT) is expected to develop at the high-density part of the probability density function of mass- ($\rho$-PDF) and column-density ($N$-PDF). The latter has been observed in a number of studies of star-forming regions \citep{Kainulainen_ea_09, Schneider_ea_15a, Schneider_ea_15b}. The PLT of the $\rho$-PDF has been found in numerical simulations \citep[e.g.][]{Klessen_00, Dib_Burkert_05, Kritsuk_Norman_Wagner_11, Collins_ea_12} and explanations of this phenomenon have been presented based on theoretical considerations \citep{Elmegreen_11, Girichidis_ea_14, Guszejnov_ea_18, Donkov_Stefanov_18, Elmegreen_18}. Almost all of these works are dedicated to the investigation of dense molecular gas, while \citet{Elmegreen_18} supposes that PLTs should also be observed at larger scales comparable with the Galactic scale height.

Gas structures, that correspond to spatial scales within the PLT range, could be transient or collapsing, depending on the local Jeans mass. It has been shown from simple hierarchical considerations that the mass function of such condensations $d N/d\log m$ should possess a slope $-1$ \citep{Fleck_96}, since each small spatial scale at the hierarchy bottom is included in each large scale at the top. Such slope is typical for fractals, whose dynamics is determined by steady state \citep{Elmegreen_Falgarone_96}. In the case one tightens the scope of consideration to collapsing structures only, a Salpeter slope $\sim-1.3$ \citet{Salpeter_55} might be expected \citep{Hennebelle_Chabrier_08}. 

In this Paper we aim to model the mass function of prestellar cores (CMF) generated in molecular clouds (MCs) which are characterized by a pure power-law $\rho$-PDF. A statistical approach is justified since both the CMF and the PDF are statistical descriptions of star-forming regions. In many regions the high-mass slope of the derived CMF turns out to be indistinguishable within the 1$\sigma$ uncertainty, due to the moderate statistics \citep[e.g.,][]{ALL_07, Reid_Wilson_06, Enoch_ea_08, Ikeda_Kitamura_09, Polychroni_ea_13}. On the other hand, the diversity of $\rho$-PDFs as testified from numerous simulations of star-forming regions is large \citep{Klessen_00, Dib_Burkert_05, Kritsuk_Norman_Wagner_11, Collins_ea_12, Federrath_Klessen_13, Girichidis_ea_14}. To deal with this issue, we divide MCs in ensembles according to the PDF slope and each ensemble is represented by a single spherical cloud. The cores are considered as elements of self-similar structure typical for fractal clouds and are modeled by spherical objects populating each cloud shell. Their mass function is derived from basic relations between core quantities whereas cores' statistics is obtained from the $\rho$-PDF.

Our approach is presented in detail in Section \ref{Sec_Setting_of_the_model}. We model the prestellar cores as abstract, homogeneous objects (called simply `cores') which obey appropriate assumptions, reflecting the basic properties of real condensations. Starting from these assumptions and from the properties of the $\rho$-PDF, we derive a mass-density relationship, which our statistical cores should obey. Then, in Section \ref{Sec_The CMF}, we derive the general CMF within this framework. The conditions for core collapse are analyzed in Section \ref{Sec_Jeans analysis} and hence the mass function of unstable cores is derived (Section \ref{Sec_CMF_ff_weigthed}). We discuss two issues on the model's applicability in Section \ref{Sec_Discussion} and conclude with a summary of the Paper (Section \ref{Sec_Summary}). To facilitate the reading of the Paper, a list of frequently used symbols and notations is provided in Table \ref{table_symbols}.

\begin{table}
\caption{Frequently used notations}
  \label{table_symbols}
%    \centering
  \begin{tabular}{lp{6.2cm}}
 \hline 
 \hline
Variable & Description \\
\hline
\multicolumn{2}{l}{{\textbf{\textit{Parameters of the entire cloud}}}}\\
$\ell_{\rm c}$ & Total size\\     
$\ell_{0}$  & Size of the homogeneous inner part\\    
$\ell$ & (Considered) scale\\     
$M_{\rm c}$ & Mass\\
$M_{\rm J,\,c}$ & Jeans mass\\
$\rho_{\rm c}$ & Density at the cloud's edge \\    
$\rho_{0}$  & Density of the inner part of the cloud \\
$\langle \rho \rangle_{\rm c}$ & Average density\\
$q$ & Slope of the $\rho$-PDF \\
$p$ & Exponent of the cloud density profile \\ 
$\gamma$ & Exponent of the mass-size relationship \\
$\kappa$ & Parameter that accounts for the fragmentation of the cloud \\
$N_{\rm c}$ & Total number of cores for the entire cloud \vspace{6pt}\\
\multicolumn{2}{l}{{\textbf{\textit{Parameters of the core population}}}}\\
$l$ & Core size \\
$m$ & Core mass \\
$l_{\rm n},~m_{\rm n},~\rho_{\rm n}$ & Normalization units of size, mass and density \\
$x$ & Exponent of the mass-density relationship (structure parameter) \\ 
$\rho_{\rm thres}$ & Threshold density for star formation \\
$M_{\rm ch}$ & Characteristic mass (lower mass limit) of the CMF \\
$\Gamma$ & Slope of the CMF \\
\hline
\end{tabular}
% \smallskip 
\end{table}

\section{Setting of the model}   
\label{Sec_Setting_of_the_model}

\subsection{Cloud model and its pdf}
\label{Subsec_Cloud_model_and_pdf}

We use the abstract model of molecular clouds that was introduced in \citet[][hereafter, DVK17]{DVK_17}. The basic properties and assumptions are summarised below. A spherical cloud with mass density profile $\rho(\ell)$ is considered. The scales $\ell_{0} \ll\ell \leq \ell_{\rm c}$ are defined simply as radii measured from the centre of the sphere to a given density level, where $\ell_{\rm c}$ is the size of the entire cloud and $\ell_{0}$ is the size of its homogeneous inner part. Thus the scales are derived from the volume-weighted $\rho$-PDF $p(s)$:
\begin{eqnarray} \label{def_abstr-scale}
\ell(s)=\ell_{\rm c}\left(\int\limits_{s}^{\infty}p(s)ds\right)^{1/3}~, 
\end{eqnarray} 
where $s=\ln(\rho/\rho_{\rm n})$ is the logarithmic density with the average density of the entire cloud $\rho_{\rm n}\equiv\langle\rho\rangle_{\rm c}$ chosen as a normalization unit. Using this definition, the size of the homogeneous inner part $\ell_0 \ll \ell$ is neglected to simplify the calculations. The upper integration limit is taken to be infinity, i.e. the density in the cloud inner part corresponds to very large densities compared to the density at the cloud's edge. Thus $\ell(s)$ is radius of the sphere, corresponding to density level $\rho$. This radius is not related to a size of any contiguous objects, delineated on MC intensity maps or through clump extraction techniques. It contains implicitly the physics of the considered MC through the $\rho$-PDF. \citet{Li_Burkert_16} use a similar definition aimed to simplify the cloud structure in their model.

\begin{figure} 
\begin{center}
% \includegraphics[width=70mm]{MC_class_of_equivalence.eps}
% \resizebox{\hsize}{!}{\includegraphics{<MC_class_of_equivalence.eps>}}
\includegraphics[width=70mm]{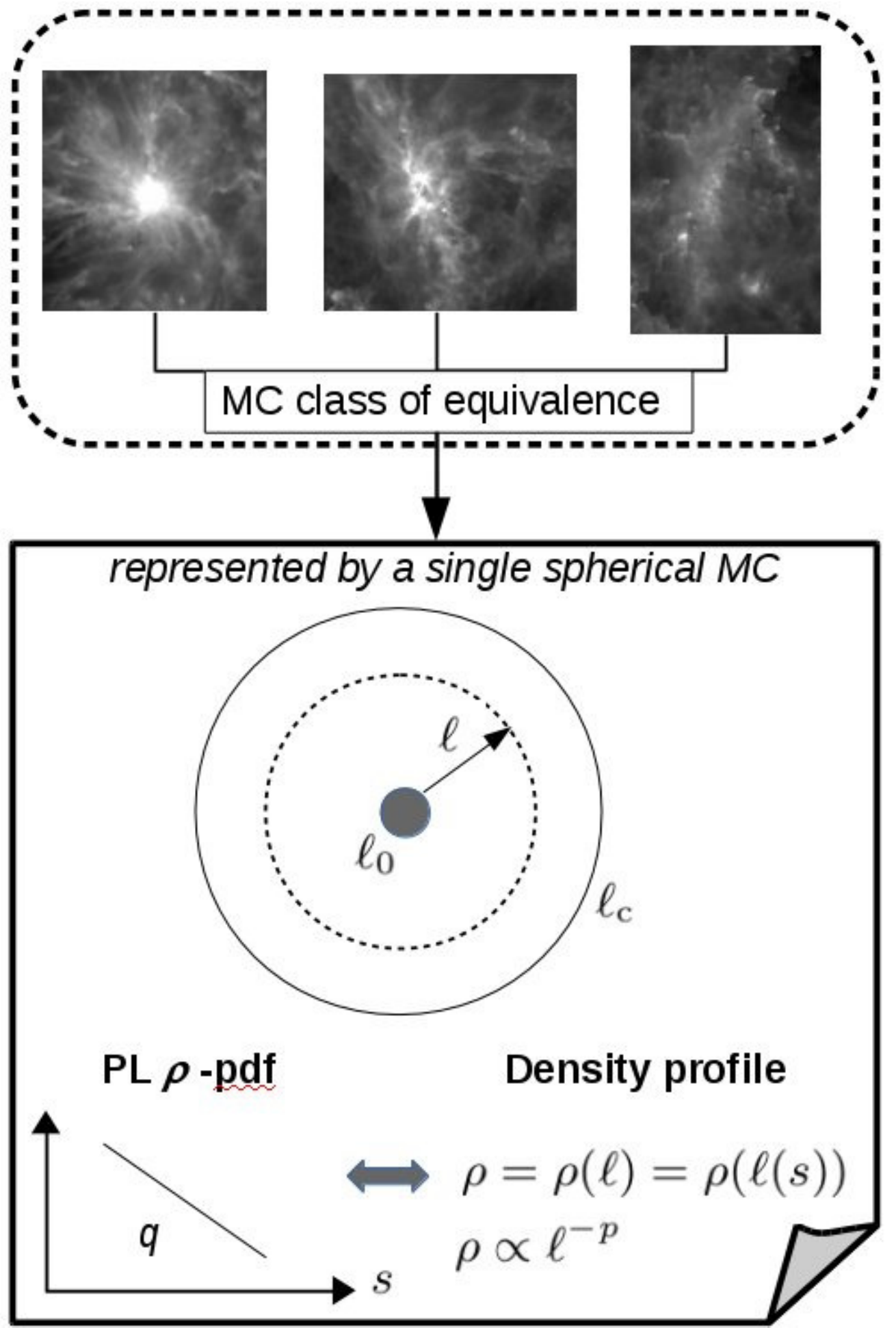}
\vspace{0.1cm}  
\caption{On the concept of the MC class of equivalence (after \citealt{DVK_17}).}
\label{fig_MC_class_of_equivalence}
\end{center}
\end{figure}

The cloud is taken as representative of the so called {\it MC class of equivalence}, introduced in DVK17. Fig. \ref{fig_MC_class_of_equivalence} schematically illustrates the concept. By assumption, all class members are characterized by single $\rho$-PDF, single cloud size ($\ell_{\rm c}$), single size ($\ell_0$) and density ($\rho_0$) of the cloud inner part  and density at the cloud's edge ($\rho_{\rm c}$). In this Paper, we add also the assumption that the cloud is isothermal, with temperature $T$.  We point out that individual class members could widely differ in their morphology and physics. The MC class of equivalence shall be conceived as a statistical ensemble. Its averaged (abstract) member possesses spherical symmetry and isotropy and is statistically representative for the behavior of any single class member. 

% To clarify the relation between abstract scale, defined through the pdf, and scales, defined/measured on the basis of some observational/numerical approach. 
In this Paper we consider a $\rho$-PDF that consists only of a PLT, with a slope $q<-1$:

\begin{eqnarray}
\label{eq_PL_pdf}
p(s)ds= A_{\rm s} \exp(qs)ds= A_{\rm s} \left(\frac{\rho}{\rho_{\rm n}}\right)^q d\ln(\rho/\rho_{\rm n})~.
\end{eqnarray}
Here $A_{\rm s}$ is the normalisation constant and can be obtained from the condition $\int_{\rho_{\rm c}}^{\rho_0} p(s)ds=1$. One gets:

\begin{eqnarray}
\label{A_s}
A_{\rm s}=\frac{q}{\exp(qs_0)-\exp(qs_{\rm c})}\approx \nonumber \\
\approx (-q)\exp(-qs_{\rm c})=(-q)\left(\frac{q}{1+q}\right)^q~,
\end{eqnarray}
making use of a formula for $\langle\rho\rangle_{\rm c}$ obtained in DVK17. On the reasonable assumption $\rho_{\rm c}\equiv\rho(\ell_{\rm c}) \ll\rho_0$, the averaged density of the entire cloud becomes a simple function of the cloud's edge density and the $\rho$-PDF slope $q$:

\begin{eqnarray}
\label{eq_rho_n-q_rho_c}
\rho_{\rm n}\equiv\langle\rho\rangle_{\rm c}=\frac{q}{1+q}\rho_{\rm c}~~.
\end{eqnarray}

\subsection{Assumptions about the core population}
\label{Subsec_Model_Postulates}

Within the presented cloud model of DVK17, now we implement assumptions about the population of prestellar cores in MCs of given class of equivalence. These include: i) relations between core mass, density and size, and, ii) a rule for statistical weighting of cores according to their densities which is needed to derive the core mass function in Section \ref{Sec_The CMF}. 

\subsubsection{Relations for statistical cores}
\label{Subsubsec_Postulates_stat_cores}
The prestellar cores are modeled by abstract statistical objects called hereafter merely ``cores''. They are homogeneous spheres with mass $m$ and size (radius) $l$. The cloud shell, corresponding to log-density range $[s,\,s+ds]$ at a given density level $\rho=\rho_{\rm n}\exp(s)$, is populated by $|dN_{\rho}|$ cores\footnote{The absolute value is to be taken since the total number of cores $N_{\rho}=\int_{\rho_0}^{\rho} dN_{\rho}$ decreases with increasing the density threshold $\rho$ (see Section \ref{Subsubsec_Postulate_core_counting}).}. 

We postulate a natural relation between core density, mass and size:
\begin{eqnarray} \label{Equation: Density-mass-size relation}
   \frac{m}{m_{\rm n}}=\frac{\rho}{\rho_{\rm n}}\left(\frac{l}{l_{\rm n}}\right)^3
\end{eqnarray}
as well a commonly adopted core mass-size relation of power-law type:
\begin{eqnarray} \label{Equation: Core mass-size relation}
   \frac{m}{m_{\rm n}}=\left(\frac{l}{l_{\rm n}}\right)^\gamma~,
\end{eqnarray}
where $m_{\rm n}$, $\rho_{\rm n}$ and $l_{\rm n}$ are normalization units. The second relation is studied in many works on core populations as the exponent $\gamma$ is often taken to be constant (independent on the scale). In the proposed model we adopt the assumption of self-similarity in turbulent fractal clouds, i.e. that the core mass-size relation should reflect the general cloud structure in terms of abstract scales: $M(\ell)\propto\ell^\gamma$. Then, in case of a purely power-law $\rho$-PDF, the mass scaling exponent is a function solely of its slope (see Sect. 3.2 in DVK17):
\begin{eqnarray}
    \gamma=3+\frac{3}{q}
\end{eqnarray}
% ,&~~~~~~~~~\gamma=&\frac{3}{1-x}=\frac{3(1+q)}{q}

\begin{figure*} 
\begin{center}
\includegraphics[width=.7\textwidth]{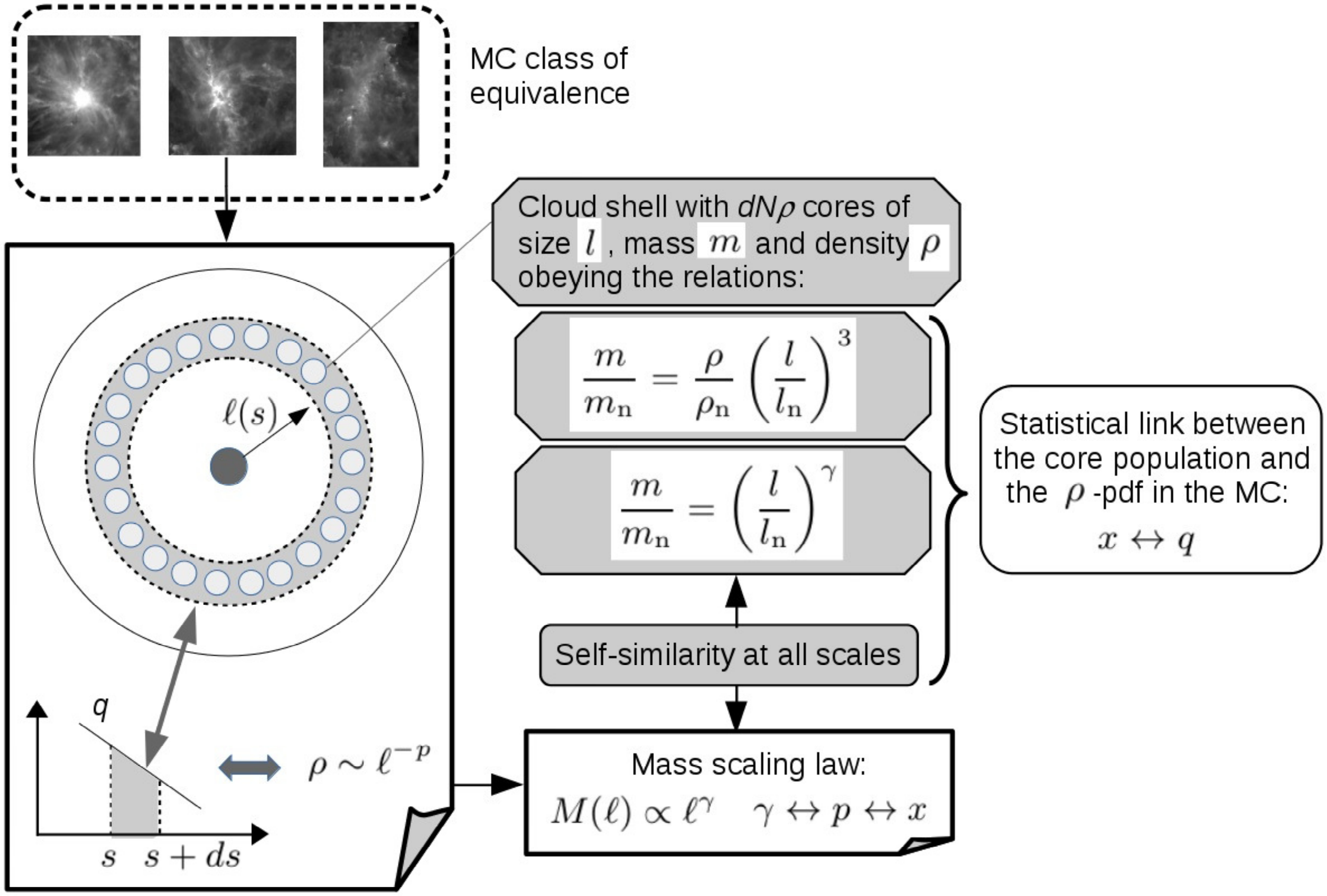}
\vspace{0.1cm}  
\caption{Sketch of the model setting which shows the statistical link between the density distribution in the cloud and the core population. The double-ended arrows denote one-to-one correspondence between quantities. The concepts introduced in this work (in regard to the DVK17 model) are put in grey frames.}
\label{fig_Statistical_link_x-q}
%  (after \citealt{DVK_17})
\end{center}
\end{figure*}

Combining the assumed relations between core quantities (Eqs. (\ref{Equation: Density-mass-size relation})-(\ref{Equation: Core mass-size relation})), one can derive the following mass-density relation for the cores:  % 
\begin{eqnarray}
\label{Equation: Mass-density relation}
   \frac{\rho}{\rho_{\rm n}}=\left(\frac{l}{l_{\rm n}}\right)^{\gamma-3}=\left(\frac{m}{m_{\rm n}}\right)^x~,
\end{eqnarray}
where the power index 
\begin{eqnarray}  
\label{Equation: Structure parameter-slope relation}
  x=\frac{3-\gamma}{\gamma}=\frac{1}{1+q}~~.
\end{eqnarray}
is called {\it structure parameter}. We point out that the existence of a power-law mass-density relation for core populations is supported by the main scenarios of core formation and/or evolution. If the cores have formed via purely turbulent fragmentation, it is derived from combination of the velocity scaling law with shock-front conditions \citep{Padoan_Nordlund_02}. In gravoturbulent scenarios, the core mass-density relation is an outcome of energy balance (virial-like relations) at different spatial scales \citep{DVK_12}. \citet{DVK_11} substantiated it theoretically, estimating the range of values of the structure parameter $x$ from equipartitions between various forms of energy in evolved MCs. Fig. \ref{fig_Statistical_link_x-q} illustrates how the basic elements of the proposed model are linked. 

Numerical estimates of the exponents in those relations will be useful for reference in the considerations hereafter. Typical slopes $-4\leq q \leq-1.5$ for evolving PLTs \citep{Kritsuk_Norman_Wagner_11, Collins_ea_11, Girichidis_ea_14} yield $-0.33\geq x \geq-2$, $2.25\geq \gamma \geq1$.

One has some freedom to choose the normalization units in the relations \ref{Equation: Density-mass-size relation} and \ref{Equation: Core mass-size relation}. A widely used choice of normalization unit of density in numerical simulations is the mean region/cube density. In this work, following DVK17, we opt for the mean cloud density (eq. 4):
\begin{eqnarray}
\label{Equation: Normalization coefficient of density}
\rho_{\rm n}\equiv\langle\rho\rangle_{\rm c}=\frac{q}{1+q}\rho_{\rm c}~.
\end{eqnarray}
Regarding the normalization unit of size, it is natural to set it to be comparable to the core sizes, i.e. in a broad range of scales below the cloud size. For simplicity we take: 
\begin{eqnarray}
\label{Equation: Normalization coefficient of size}
l_{\rm n}\equiv\kappa\ell_{\rm c}~,~~~0<\kappa\leq 1~,&~
\end{eqnarray}
where $\kappa={\rm const}(\ell)$ is a model parameter. The meaning of $\kappa$ will be clarified further throughout Sections \ref{Sec_The CMF}-\ref{Sec_CMF_ff_weigthed}. The normalization unit of mass $m_{\rm n}$ is obtained from the condition of mass conservation at a given scale $\ell$ (see Section \ref{Subsubsec_Postulate_core_counting}).  

%Also we postulate that $x_{\rm local} \equiv x_{\rm global}$ and finally we have: $x_{\rm global}=1/(1+q)$. This is an obvious suggestion because in the case of PL-pdf $x=const(\ell)$ (DVK17).

By use of the normalization units, we define logarithmic variables for mass, size and volume as follows:
\begin{eqnarray}
\label{s_m,s_l}
s_{\rm m}\equiv\ln(m/m_{\rm n}),~~s_{\rm l}\equiv\ln(l/l_{\rm n}),~~s_{\rm v}\equiv\ln(v/v_{\rm n})~~,
\end{eqnarray}
where $v=(4\pi/3)l^3$ is the core volume and $v_{\rm n}=(4\pi/3)l_{\rm n}^3$ is the volume normalization unit. Now one is able to derive from the density PDF (Eq. (\ref{eq_PL_pdf})) statistical distributions of core masses, sizes and volumes which are power-law functions with exponents depending on $q$ and on the structure parameter $x$. This is done in Appendix \ref{Appendix: Statistical distributions}. We note here only that
\begin{eqnarray}
 \label{Equation: Correspondence between the pdfs}
 p(s)ds=p(s_{\rm m})ds_{\rm m}=p(s_{\rm l})ds_{\rm l}=p(s_{\rm v})ds_{\rm v}~,
\end{eqnarray}
due to the one-to-one correspondence between the density, mass, size and volume of the cores.

\subsubsection{Weighting of cores according to their density}
\label{Subsubsec_Postulate_core_counting}

The key issue in our model is how to weight the contributions of statistical cores with different densities to the total statistics. If $V_{\rm c}=(4/3)\pi\ell_{\rm c}^3$ is the total volume of the cloud, we postulate that the volume $dV_s=-V_{\rm c}p(s)ds$ of the shell, corresponding log-density range $[s,s+ds]$, is equal to the sum of volumes of the cores it contains. Hence, if $dN_\rho$ is the contribution of a shell with density $\rho$ to the total number of cores, we get:
\begin{eqnarray}
\label{postulate_dN_rho}
\frac{4\pi}{3}l^3dN_\rho\equiv - V_{\rm c}p(s)ds~~\Leftrightarrow~~l^3dN_\rho\equiv -\ell_{\rm c}^3p(s)ds~.
\end{eqnarray}
Note that $dN_{\rho}$ is negative, since the total number of cores $N_{\rho}$ down to given density threshold $\rho$ decreases as the density increases up to $\rho_0$. The same holds for the volume $dV_s$. We note that $V_s\equiv V_{\ell}$.

One must check whether the volume and mass are conserved at a given scale $\ell_{\rm c}\geq\ell\gg\ell_0$ (i.e. at given density threshold $\rho_{\rm c}\leq\rho\ll\rho_0$). The volume of a scale $\ell$ is calculated straightforward:
$$V_\ell=\frac{4\pi}{3}\ell^3~~.$$
To calculate the mass of the scale $M_{\ell}$, we invoke the scaling relations for mass and density profile for a given scale in case of power-law pdf, derived in DVK17,

\begin{eqnarray}
\label{eq_M_ell-rho_ell}
M_{\ell}=M_{\rm c}\left(\frac{\ell}{\ell_{\rm c}}\right)^\gamma~, \nonumber \\
\rho=\rho_{\rm c}\left(\frac{\ell}{\ell_{\rm c}}\right)^{-p}~,
\end{eqnarray}
where $p$ is the density profile exponent and $M_{\rm c}=(4\pi/3)\ell_{\rm c}^3 \langle\rho\rangle_{\rm c}$ is the total mass of the cloud. A simple relation between $p$ and the slope of the high-density power-law part of the $\rho$-PDF in spherically symmetric clouds has been derived analytically: $p=-3/q$ \citep{Federrath_Klessen_13, Girichidis_ea_14}. By use of it and combining Eqs. (\ref{eq_rho_n-q_rho_c}), (\ref{Equation: Structure parameter-slope relation}) and (\ref{eq_M_ell-rho_ell}), one gets:
\begin{eqnarray}
\label{eq_M_ell-M_c}
% \begin{aligned}
M_\ell &=&M_{\rm c}\left(\frac{\ell}{\ell_{\rm c}}\right)^\gamma=M_{\rm c}\left(\frac{\rho}{\rho_{\rm c}}\right)^{-\gamma/p}= \nonumber\\
~&=&M_{\rm c}\left(\frac{\rho}{\rho_{\rm n}}\right)^{1+q}\left(\frac{q}{1+q}\right)^{1+q}~~.
%\end{aligned}
\end{eqnarray}

On the other hand, the volume and the mass of a given scale $\ell$ can be calculated as sums of the volumes and masses of the cores populating this scale. Taking into account that $V_0\ll V_{\ell}$ and using Eq. (\ref{postulate_dN_rho}), one obtains for the volume:
$$\sum_{\rho}^{\rho_0} \frac{4\pi}{3}l^3 |\Delta N_{\rho}|= \int_{\rho}^{\rho_0}-\frac{4\pi}{3}l^3 dN_{\rho}=$$
$$=\int_{\rho}^{\rho_0} V_{\rm c}p(s)ds=\int_{\rho}^{\rho_0} -dV_s \simeq V_{\ell}~.$$
That simply means the volume conservation is trivial, since the cores are accounted for in Eq. (\ref{postulate_dN_rho}) through their volumes. For the mass one obtains accordingly, in view of the model setup $\rho_{\rm c}\leq\rho\ll\rho_0$:
\begin{eqnarray}
\begin{aligned}
\sum_{\rho}^{\rho_0} m |\Delta N_{\rho}|&= \int_{\rho}^{\rho_0} m l^{-3} \ell_{\rm c}^3 p(s)ds~ \nonumber\\
&=\frac{m_{\rm n}}{l_{\rm n}^3} \ell_{\rm c}^3 \int_{\rho}^{\rho_0} \frac{m}{m_{\rm n}} \left(\frac{l}{l_{\rm n}}\right)^{-3} p(s)ds \nonumber\\
&=\frac{m_{\rm n}}{l_{\rm n}^3} \ell_{\rm c}^3 \int_{\rho}^{\rho_0} \left(\frac{\rho}{\rho_{\rm n}}\right) p(s)ds \nonumber \\
&\simeq m_{\rm n} \left(\frac{\ell_{\rm c}}{l_{\rm n}}\right)^3 \left(\frac{q}{1+q}\right)^{1+q}\left(\frac{\rho}{\rho_{\rm n}}\right)^{1+q} \nonumber \\
&=\frac{m_{\rm n}}{\kappa^3}\left(\frac{q}{1+q}\right)^{1+q}\left(\frac{\rho}{\rho_{\rm n}}\right)^{1+q}~~. \nonumber
\end{aligned}
\end{eqnarray}

Equating the above expression for $M_{\ell}$ and formula (\ref{eq_M_ell-M_c}), we derive an expression for the mass normalization unit $m_{\rm n}$ which links it to the parameter $\kappa$ (Eq. (\ref{Equation: Normalization coefficient of size})):
\begin{eqnarray}
\label{eq_m_n}
m_{\rm n}=\kappa^3 M_{\rm c}~~.
\end{eqnarray}
This formula leads (in view of Eq. (\ref{Equation: Density-mass-size relation})) to an important relation for the cores:
\begin{eqnarray}
\label{eq_m_rho_l}
\frac{m}{\rho v}=\frac{m_{\rm n}}{\rho_{\rm n} v_{\rm n}}=1~.
\end{eqnarray}
Some statistical quantities of the cores and their relations are derived in the Appendix \ref{Appendix: Averaged quantities} by use of the formulae obtained in Section \ref{Subsec_Model_Postulates}.

Now one is able to calculate two quantities which are measures of the total number of cores above a given density level ($N_{\rho}$) and in the entire cloud ($N_{\rm c}$):
\begin{eqnarray}
\label{eq_N_rho}
\begin{aligned}
N_{\rho}&= \int\limits_{\rho_0}^{\rho} dN_{\rho}= \int\limits_{\rho}^{\rho_0} \frac{1}{\kappa^3} \left(\frac{l}{l_{\rm n}}\right)^{-3} p(s)ds= ... \\
&= \left(\frac{-q}{\kappa^3}\right) \left(\frac{q}{1+q}\right)^q \ln(\rho_0/\rho)~.
\end{aligned}
\end{eqnarray}
\begin{eqnarray}
\label{eq_N_c}
\begin{aligned}
N_{\rm c}\equiv N_{\rho_{\rm c}}=-\frac{q}{\kappa^3}\left(\frac{q}{1+q}\right)^q\ln(\rho_0/\rho_{\rm c})~~.~~~~
\end{aligned}
\end{eqnarray}
The latter quantity is useful for assessment of the conditions for core collapse (Section \ref{Sec_Jeans analysis}). 

\section{The Core Mass Function}
\label{Sec_The CMF}

In this Section we are going to derive the Core Mass Function (hereafter CMF) within the framework of the presented model. The statistical contribution of cores in a given shell $[s,s+ds]$ has been calculated in Section \ref{Subsubsec_Postulate_core_counting} (Eq. (\ref{postulate_dN_rho})): $dN_{\rho}= -(\ell_{\rm c}/l)^3 p(s)ds$. Due to the one-to-one correspondence between core density and core mass, we have $dN_{\rm m}=dN_{\rho}$ and $p(s)ds=p(s_{\rm m})ds_{\rm m}$ (cf. Eq. (\ref{Equation: Correspondence between the pdfs})). Hence:
$$\frac{dN_{\rm m}}{d\ln\left(\frac{m}{m_{\rm n}}\right)}= -\left(\frac{\ell_{\rm c}}{l}\right)^3 p(s_{\rm m})~~.$$

After some algebraic operations and by use of Eqs. (\ref{Equation: Normalization coefficient of size}) and (\ref{Equation: Structure parameter-slope relation}) one gets: 
$$(\ell_{\rm c}/l)^3=(\ell_{\rm c}/l_{\rm n})^3 (l_{\rm n}/l)^3=(1/\kappa)^3(m/m_{\rm n})^{-3/\gamma}~.$$
Now, replacing $p(s_{\rm m})$ with $A_{\rm m}(m/m_{\rm n})^{xq}$ from Eq. (\ref{eq_mass_distribution}) and in view of the relation $-3/\gamma+xq=0$ (cf. Eq. (\ref{Equation: Core mass-size relation})), we obtain a differential core mass distribution:
\begin{eqnarray}
\label{cores_CMF}
\frac{dN_{\rm m}}{d\ln\left(\frac{m}{m_{\rm n}}\right)}=\frac{1}{\kappa^3}\left(\frac{q}{1+q}\right)^{1+q}\left(\frac{m}{m_{\rm n}}\right)^0~.
\end{eqnarray}

This is still not a formula for the CMF since one has to take into account the fractal structure of the cloud. In other words, the r.h.s. of Eq. (\ref{cores_CMF}) must be weighted with respect to the number of scales at density level $\rho$ which are contained in the entire MC. This weighting corresponds to the physical picture of a steady state in the cloud as the material is accreted through the cloud boundary and is transferred downwards through all scales. Then, for a given scale $\ell$ with mass $M_{\ell}$, the weighting coefficient should be $M_{\rm c}/M_{\ell}$. Making use of Eqs. (\ref{eq_M_ell-rho_ell}) and (\ref{Equation: Mass-density relation}) - (\ref{Equation: Normalization coefficient of density}), one obtains
$$\frac{M_{\rm c}}{M_{\ell}}=\left(\frac{\ell}{\ell_{\rm c}}\right)^{-\gamma}=\left(\frac{\rho}{\rho_{\rm c}}\right)^{\gamma/p}=\left(\frac{q}{1+q}\right)^{\gamma/p}\left(\frac{m}{m_{\rm n}}\right)^{x\gamma/p}~,$$
and for the exponents: $\gamma/p=-(1+q)$ and $x\gamma/p=-1$. This yields a formula for the CMF:
\begin{eqnarray}
\label{eq_CMF_scale_weighted}
{\rm CMF}=\frac{1}{\kappa^3}\left(\frac{m}{m_{\rm n}}\right)^{-1}=\frac{M_{\rm c}}{M_{\odot}}\left(\frac{m}{M_{\odot}}\right)^{-1}~~,
\end{eqnarray}
where the conversion to solar units is made using Eq. (\ref{eq_m_n}): $(m/m_{\rm n})^{-1}=\kappa^3 (M_{\rm c}/M_{\odot}) (m/M_{\odot})^{-1}$. 

The derived CMF does not depend on $\kappa$ (formula \ref{eq_CMF_scale_weighted}), i.e. the latter behaves as a free parameter {\it as long one considers the cores as substructures in a fractal (self-similar) cloud}. Indeed, this result recovers the mass spectrum in the interstellar medium modeled as a scale-invariant hierarchy of density fluctuations \citep{Fleck_96}. As shown by \citet[][see Sect. 4 there]{Elmegreen_Falgarone_96}, the slope $-1$ is to be expected considering a large sample of clouds -- the fractal dimension of the whole ensemble equals the mass-size exponent $\gamma$ which yields a CMF independent on the physical conditions in an individual cloud. The construction of statistical ensemble proposed in DVK17 and in this work is consistent with their conclusion. From the point of view of observations, the total clump population in star-forming regions, extracted by use of various clump-finding techniques, display shallower or similar slopes if the CMF is fitted by {\it one} power-law function \citep{Heithausen_ea_98, Kramer_ea_98, Li_ea_07, Pekruhl_ea_13}. On the other hand, if the CMF is fitted by two power-law functions, the slope of the high-mass part is comparable or steeper than that of the stellar IMF (see \citealt{VDK_13} and references therein). Some numerical simulations \citep[e.g.][]{Dib_ea_08a, Dib_ea_08b} indicate also that the slope should steepen when high-density cores are selected. 

% \section{Jeans analyzes}
\section{Conditions for core collapse}
\label{Sec_Jeans analysis}

\begin{figure} 
\begin{center}
\includegraphics[width=84mm]{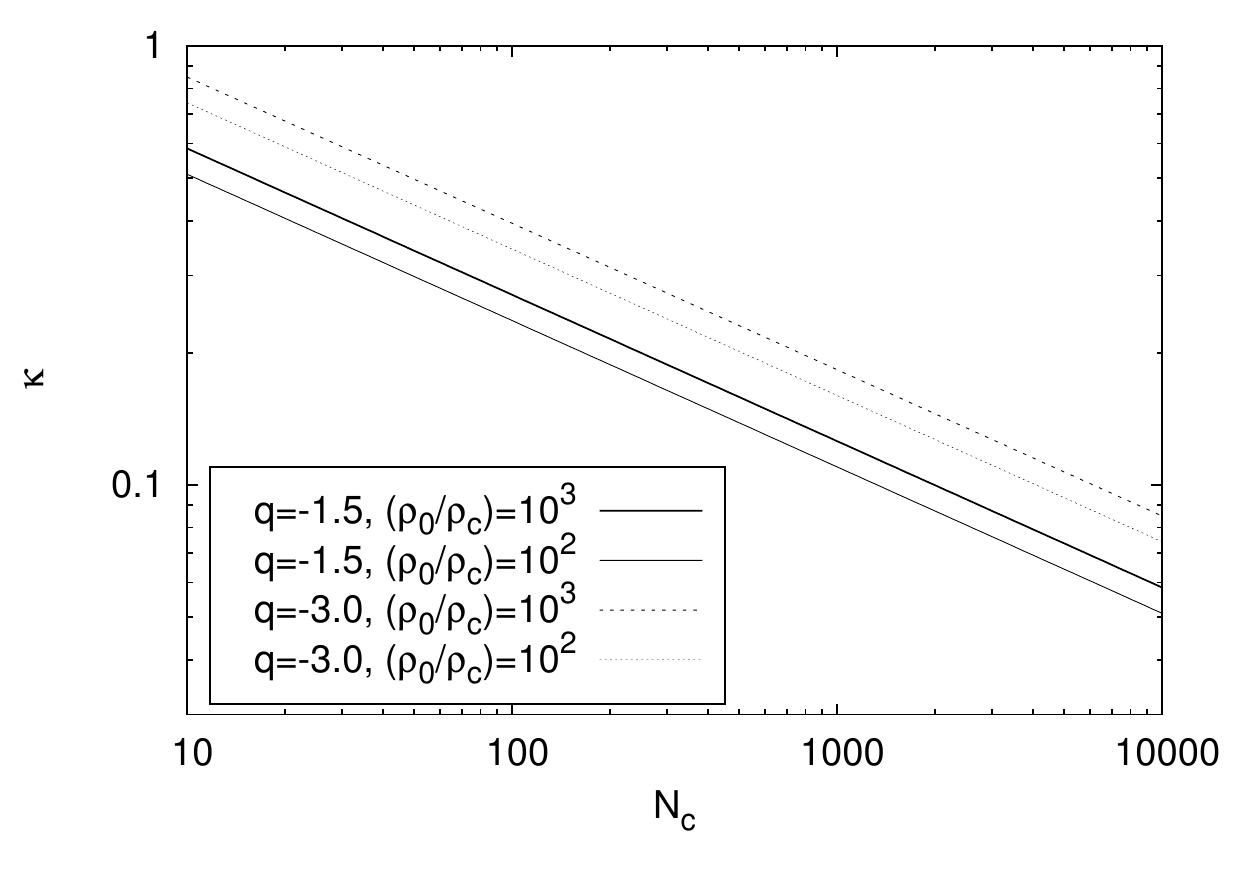}
\vspace{0.1cm}  
\caption{Relationship between the parameter $\kappa$ and the total number of detected cores (Eq. (\ref{eq_kapa-N_c})), for different choice of the $\rho$-PDF slope and density contrast.}
\label{fig_kappa_fid}
\end{center}
\end{figure}

Now let us analyze the ability of cores of given density $\rho$ to collapse. We introduce the Jeans mass at density $\rho$ in the form
\begin{equation}
\label{Jeans mass}
m_{\rm J}(\rho)= B_{\rm T}\rho^{-1/2}~~,
\end{equation}
where $B_{\rm T}= 1.22 c_{\rm s}^{3}/G^{3/2}\propto T^{3/2}= {\rm const}$, due to the assumption of isothermality of the  cloud. Recalling the derived core mass-density relationship (Eq. (\ref{Equation: Mass-density relation})), one gets for the mass of cores at density level $\rho$: $m(\rho)= m_{\rm n}(\rho/\rho_{\rm n})^{1/x}= \kappa^3 M_{\rm c} (\rho/\rho_{\rm n})^{1+q}$. This yields for the ratio of the core mass to the local Jeans mass
\begin{eqnarray}
\frac{m(\rho)}{m_{\rm J}(\rho)}= \frac{\kappa^3 M_{\rm c} \rho_{\rm n}^{1/2}}{B_{\rm T}} \left(\frac{\rho}{\rho_{\rm n}}\right)^{q+3/2}~,\nonumber
\end{eqnarray}
which is transformed by use of Eq. (\ref{Equation: Normalization coefficient of size}) to:
\begin{eqnarray}
\label{eq_m/m_J}
\frac{m(\rho)}{m_{\rm J}(\rho)}= \frac{\kappa^3 M_{\rm c} \langle\rho\rangle_{\rm c}^{1/2}}{B_{\rm T}} \left(\frac{\rho}{\rho_{\rm n}}\right)^{q+3/2}\!\!\!\!\!\!= \kappa^3 \frac{M_{\rm c}}{M_{\rm J,\,c}} \left(\frac{\rho}{\rho_{\rm n}}\right)^{q+3/2}\!\!\!\!\!,
\end{eqnarray}
where $M_{\rm J,\,c}=B_{\rm T}\langle\rho\rangle_{\rm c}^{-1/2}$ is the Jeans mass for the entire cloud. The coefficient $(\kappa^3 M_{\rm c}/M_{\rm J,\,c})$ of this power-law relationship is a function only of the {\it global cloud parameters}. In particular, the parameter $\kappa$ (formula (\ref{Equation: Normalization coefficient of size})) is related through Eq. (\ref{eq_N_c}) to the total number of cores and the characteristics of the $\rho$-PDF:
\begin{eqnarray}
\label{eq_kapa-N_c}
\kappa=\frac{f(q,\rho_0/\rho_{\rm c})}{N_c^{1/3}}.
\end{eqnarray}
For typical PDF slopes $-3\le q \le -1.5$ and density contrasts $10^2\lesssim (\rho_0/\rho_{\rm c})\lesssim 10^3$ (corresponding to a well resolved PLT in simulations), the numerator in the expression above is a slightly varying function with values between 1 and 2. Thus, $\kappa^3$ reflects mainly the total number of cores and its reciprocal quantity $1/\kappa^3$ can be interpreted as a measure of how fragmented the cloud is (`index of MC fragmentation'). An appropriate (constant) value of $\kappa$ is to be determined from $N_{\rm c}$  (Fig. \ref{fig_kappa_fid}).

When considering the gravitational fragmentation of the cloud, as opposed to simply fractal density distributions, then $\kappa$ is no longer a free parameter because additional physical processes, such as the competition between gravitational collapse and thermal pressure, play a role. As shown in the next Section, the gravitational instability modifies the construction of the CMF.

\begin{figure*} 
\begin{center}
\includegraphics[width=1.\textwidth]{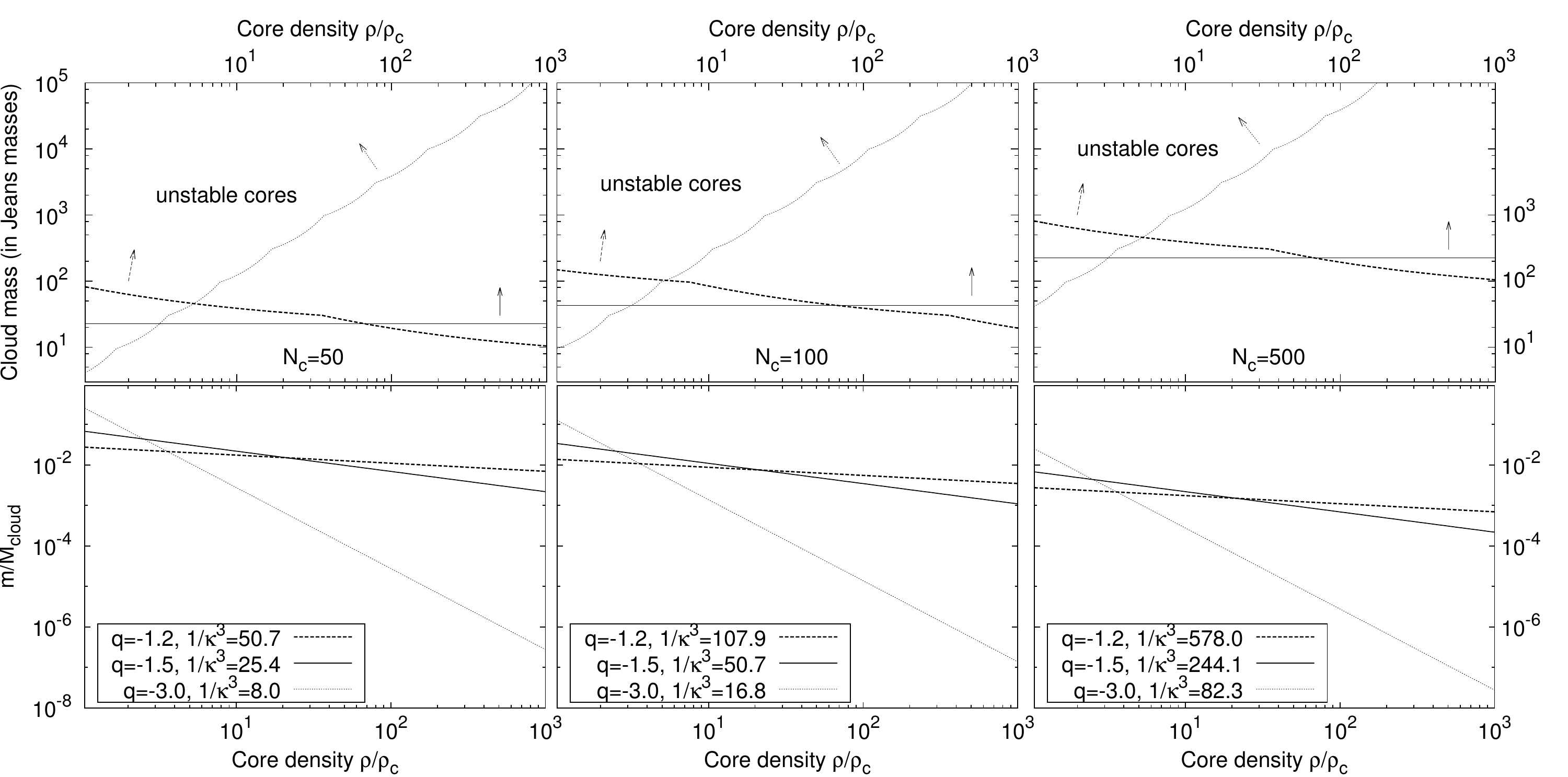}
\vspace{0.2cm}  
\caption{On the conditions for core collapse, for different total number of cores (columns) and PDF slopes. {\it Top:} The function $m/m_{\rm J}$ (Eq. (\ref{eq_m/m_J})) projected on the plain ``core density vs. cloud mass''. Lines denote the condition $m/m_{\rm J}=1$; the domains of unstable cores for the three exemplary values of $q$ are shown with arrows. An isothermal Jeans mass within the cloud $M_{\rm J,\,c}$ for $T=30$~K is adopted. {\it Bottom:} Mass-density relationship for cores (Eq. (\ref{Equation: Mass-density relation})) where the mass is given in units of the total mass of the cloud. The density contrast in the cloud is fixed at $(\rho_0/\rho_{\rm c})=10^{3}$.}
\label{fig_core_collapse_conditions}
\end{center}
\end{figure*}

Setting $m/m_{\rm J}\ge 1$ as a condition for core collapse, one arrives from Eq. (\ref{eq_m/m_J}) at three different scenarios for star formation in the cloud. They are illustrated in Fig. \ref{fig_core_collapse_conditions} where $\kappa$ is calculated from Eq. (\ref{eq_kapa-N_c}) for the given slope and $N_{\rm c}$: 
\begin{itemize}
\item $q+3/2=0~~\Leftrightarrow~~q=-3/2$ \vspace{6pt}\\
This slope corresponds to a well developed PLT at an advanced evolutionary stage in self-gravitating media \citep{Kritsuk_Norman_Wagner_11, Girichidis_ea_14}. In this special case evidently {\it all cores in the cloud will be either gravitationally stable or unstable}, depending on the Jeans content of the cloud $M_{\rm c}/M_{\rm J,\,c}$ and on the index of MC fragmentation $1/\kappa^3$: 
$$m/m_{\rm J}=\kappa^3 M_{\rm c}/M_{\rm J,\,c}~~.$$ 
In a sub-Jeans cloud all cores will dissolve without any star formation. The increase of fragmentation in super-Jeans clouds is unfavourable for core collapse. Low index of fragmentation ($1/\kappa^3=25.4$) would lead to the prediction of ubiquitous core collapse even in a moderately super-Jeans clouds (solid line in Fig. \ref{fig_core_collapse_conditions}, top left). Vice versa, in the case of highly fragmented MCs with $1/\kappa^3=244.1$ (Fig. \ref{fig_core_collapse_conditions}, top right) an ubiquitous core collapse would take place only if the Jeans content is very high (say, $M_{\rm c}\gtrsim 10^2 M_{\rm J,\,c}$).

This case is an illustrative example of how the fragmentation of the cloud determines the effectiveness of local collapse. Of course, not all possible values of $\kappa$ would have physical meaning; Fig. \ref{fig_core_collapse_conditions} is intented to present simply the total general picture. Comparison of the model with samples of real clouds can impose constraints on the index of fragmentation.
\\ 

\item $q+3/2>0~~\Leftrightarrow~~-1 > q>-3/2$ \vspace{6pt}\\
In a given cloud with some fixed index of fragmentation, the ratio $m/m_{\rm J}$ increases with the core density $\rho$. This is the case of {\it a threshold core density for star formation}  $\rho_{\rm thres}$ -- all cores with $\rho\geq\rho_{\rm thres}$ will collapse. The more Jeans masses are contained in the cloud, the lower is the threshold (thick dashed line in Fig. \ref{fig_core_collapse_conditions}, top). However, shallow slopes $-1>q>-3/2$ appear rarely in simulated self-gravitating clouds; usually for restricted time spans at their late evolutionary stages \citep{Veltchev_ea_19}.  \\

\item $q+3/2<0~~\Leftrightarrow~~q<-3/2$ \vspace{6pt}\\
Steeper $\rho$-PDF slopes are typical at earlier stages of self-gravitating media as testified from $\rho$-PDFs derived from numerical simulations \citep{Kritsuk_Norman_Wagner_11, Collins_ea_12, Veltchev_ea_19}. A PLT with $q\lesssim -3$ is hardly distinguishable from the wing of a lognormal distribution. As seen in Fig. \ref{fig_core_collapse_conditions} (top, thin dashed lines), there is an {\it upper} threshold density $\rho_{\rm thres,\,up}$ in this case, for given cloud and some fixed index of fragmentation, i.e. only cores with $\rho\le\rho_{\rm thres,\,up}$ will collapse. This result is counter-intuitive at first glance but it stems from the obtained core mass-density relationship $m\propto \rho^{1/x}$ (Eq. (\ref{Equation: Mass-density relation})). More massive cores are less dense (Fig. \ref{fig_core_collapse_conditions}, bottom) and, since $x>-2$ for $q<-3/2$, core mass grows faster with decreasing density than the local Jeans mass does. 
\end{itemize}

\section{Mass function of unstable cores}
\label{Sec_CMF_ff_weigthed}
A natural assumption for dynamically evolving clouds is that the formed cores are constantly replenished. On this assumption, \citet{CKB_07} argue that the core mass function should be weighted by a coefficient accounting for dynamics of unstable cores:
$$  \left(\frac{\tau_{\rm ff}(\rho)}{\tau_{\rm ff}(\langle\rho\rangle_{\rm c})}\right)^{-1}=\left( \frac{\langle\rho\rangle_{\rm c}}{\rho}\right)^{-1/2}=\left(\frac{m}{m_{\rm n}}\right)^{x/2}~~, $$
where $\tau_{\rm ff}(\rho)$ and $\tau_{\rm ff}(\langle\rho\rangle_{\rm c})$ are the free-fall times for a core of density $\rho$ and for the entire MC, correspondingly. Those authors consider a CMF which is a combination of two power-law functions while in our statistical framework the CMF resulting from fractal hierarchical structure is a single power law (Eq. (\ref{eq_CMF_scale_weighted})). We assume that only unstable cores, fitting the criteria derived in the previous Section, will eventually collapse. To obtain their CMF, one should apply free-fall times weighting to the r.h.s. of Eq. (\ref{eq_CMF_scale_weighted}) and gets:
\begin{eqnarray}
\label{eq_CMF_collapsing_cores}
\begin{aligned}
{\rm CMF}_{\rm \tau}&=\frac{1}{\kappa^3}\left(\frac{m}{m_{\rm n}}\right)^{-1+x/2} \\
&=\kappa^{-3x/2}\left(\frac{M_{\rm c}}{M_{\odot}}\right)^{1-x/2}\left(\frac{m}{M_{\odot}}\right)^{-1+x/2}~.
\end{aligned}
\end{eqnarray}
PDF slopes of $-4\le q\le -1.5$ yield CMF slopes $\Gamma=-1+x/2$ in the range $-1.17$ and $-2$ (cf. Eq. (\ref{Equation: Structure parameter-slope relation})). Interestingly, the latter includes the classical Salpeter value $-1.33$ of the stellar IMF. Similar CMF slopes have been found in a number of observational works (Table \ref{table_CMF_parameters}). We point out as well that the slope $\Gamma$ depends implicitly on time, through the slope of the $\rho$-PDF (Eq. (\ref{Equation: Structure parameter-slope relation})). The latter is expected to get shallower in evolving self-gravitating clouds \citep{Girichidis_ea_14} which would lead to steepening of the CMF of unstable cores.

In the case $q=-3/2$ and given that all cores are unstable (see the comment in the previous Section), one obtains a steep CMF with $\Gamma=-2$. The cases with $q< -3/2$ correspond to the typical PDFs in evolving self-gravitating clouds. The upper core density threshold $\rho_{\rm thres,\,up}$ corresponds to a minimal mass of collapsing cores which serves as the characteristic mass $M_{\rm ch}$ separating the regimes of non-collapsing and collapsing cores. In that way, the model predicts a CMF in evolving self-gravitating clouds that is a combination of two power laws with slopes:
\begin{eqnarray}
\label{eq_CMF_evolving_clouds}
\Gamma=-1~,&~&~m<M_{\rm ch} \nonumber \\
\Gamma=-1+\frac{x}{2}~,&~&~
m\ge M_{\rm ch}~.\end{eqnarray}
Note that the characteristic mass $M_{\rm ch}$ is to be calculated from the condition $m (\rho)\ge m_{\rm J}(\rho)$ (Eq. (\ref{eq_m/m_J})). The coefficient in the latter can be transformed, by use of Eq. (\ref{eq_kapa-N_c}), to:
\begin{equation} \label{eq_Characteristic_mass}
f(q,\rho_0/\rho_{\rm c})^3 \frac{(M_{\rm c}/M_{\rm J,\,c})}{N_c} \sim \frac{(M_{\rm c}/M_{\rm J,\,c})}{N_c}~~.
\end{equation}

\begin{figure} 
\begin{center}
\includegraphics[width=84mm]{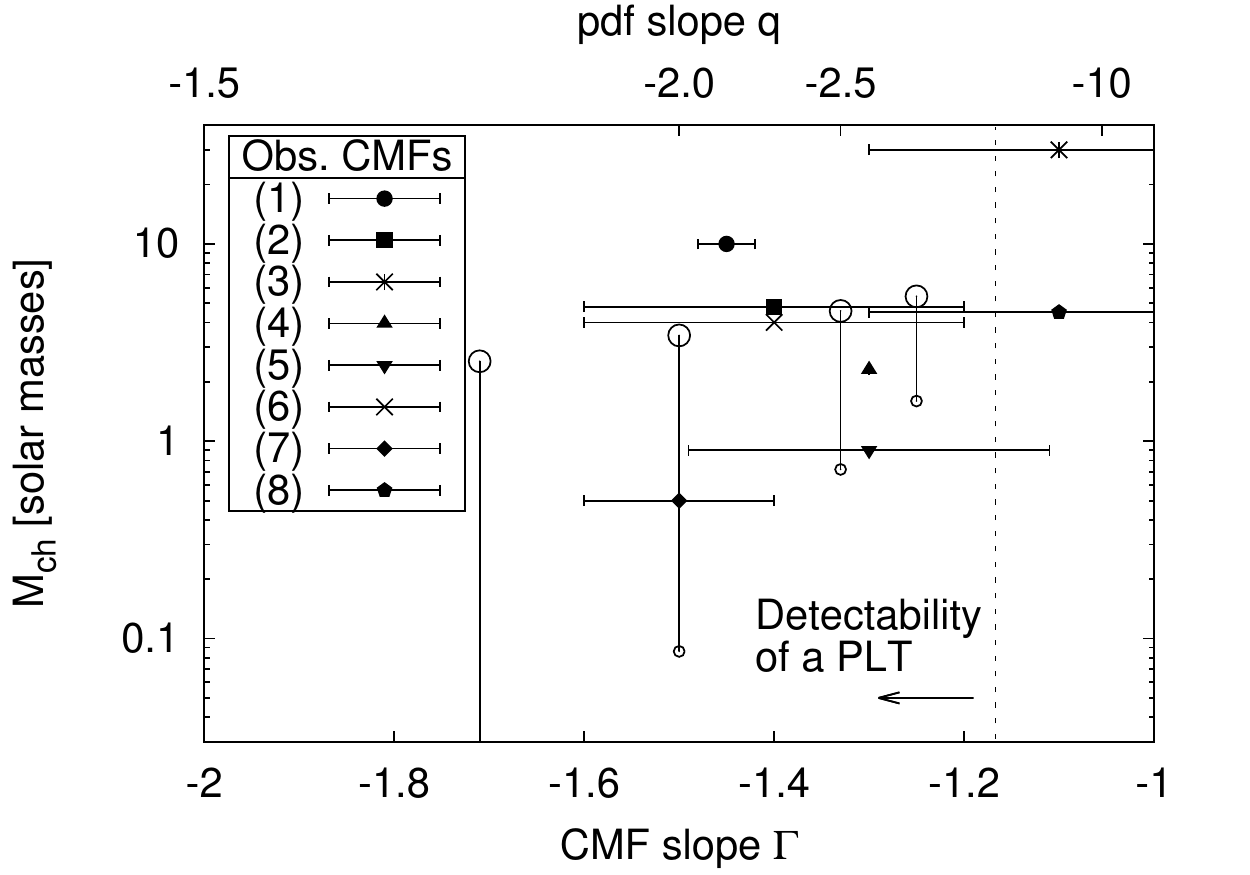}
\vspace{0.2cm}  
\caption{Parameters of modelled mass functions of unstable cores (open symbols), compared with those of the high-mass part of observational CMFs (filled symbols; the numbers correspond to the data in Table \ref{table_CMF_parameters}). Vertical solid lines denote the range of characteristic masses generated by varying $(M_{\rm c}/M_{\rm J,\,c})/N_{\rm c}$ from $0.5$ (large symbols) to $20$.0 (small symbols). The domain of PLTs which can be distinguished from a lognormal wing ($q\gtrsim-4$) is shown with vertical dashed line and an arrow. See text.}
\label{fig_CMF_parameters}
\end{center}
\end{figure}

Thus the modelled characteristic mass depends on global cloud parameters: the PDF slope $q$ (through the structure parameter $x$) and the cloud's Jeans content per core. The larger the latter quantity for a fixed $q$, the less is $M_{\rm ch}$. Note that the cloud's Jeans content per core could also take values below unity since $M_{\rm J,\,c}$ may exceed local Jeans masses substantially. Comparison of formulae (\ref{eq_kapa-N_c}), (\ref{eq_CMF_collapsing_cores}) and (\ref{eq_Characteristic_mass}) shows how $\kappa$ affects the characteristic mass. This could give additional opportunities to constrain the values of parameter $\kappa$ from observational and/or numerical studies.

In Fig. \ref{fig_CMF_parameters} we compare the CMF of unstable cores from our model with high-mass CMFs from observations of clumps in several star-forming regions (Table \ref{table_CMF_parameters}). In all but one of those works most or all of the sampled cores have been assessed as gravitationally bound. For plausible values of the Jeans content per core $0.5 \le (M_{\rm c}/M_{\rm J,\,c})/N_{\rm c} \le 20$, we find good consistency with CMFs derived from dust-extinction and dust-emission studies, with slopes $\Gamma$ close to the Salpeter value. In those cases the modelled $M_{\rm ch}$ is constrained within an order-of-magnitude range, with mean values of about $2-3$ solar masses.   

\begin{table*}
\caption{Parameters of high-mass parts of observational CMFs in some SF regions, used for comparison with modelled CMFs in Fig. \ref{fig_CMF_parameters}. All estimates are taken from the corresponding bibliographic source.}
\label{table_CMF_parameters}
\begin{center}
\begin{tabular}{lcccclc}
\hline 
\hline
SF region & Ref. & $\Gamma$ & $M_{\rm ch}$ & Tracer &  Core sizes  & Bound?\\
~         &  \#  &   ~      & [ $M_\odot$ ]&    ~    &   [ pc ]     &  ~ \\
\hline
Sample & 1  & $-1.45$ & 10.0  & C$^{18}$O & 0.08-0.45  & most? \\     
OMC-1  & 2  & $-1.40$ &  4.8  & C$^{18}$O &   $\sim 0.2$ &  yes  \\    
S 140  & 3  & $-1.10$ & 30.0  & C$^{18}$O & 0.2-0.6    & yes    \\   
Pipe   & 4  & $-1.30$ &  2.3  & dust extinction & 0.1-0.4 & ?   \\ 
Perseus, Serpens \& Oph & 5 & $-1.30$ & 0.9 & dust emission & $\sim 0.08$ & most   \\ 
Orion A  & 6 & $-1.40$ &  4.0 & dust emission & $0.03-0.09$ & yes   \\
$\rho$ Oph & 7 & $-1.50$ & 0.5 & dust emission & $\lesssim 0.01 - 0.1$ & most  \\
Vela-C & 8  & $-1.10$ & 5.0 & dust emission & $0.03-0.3$ & yes  \\
\hline
\end{tabular}
\end{center}
\smallskip
References:\\
(1) \citet{Tachihara_ea_02}; (2) \citet{Ikeda_Kitamura_09}; (3) \citet{Ikeda_Kitamura_11}; (4) \citet{ALL_07}; (5) \citet{Enoch_ea_08}; (6) \citet{Polychroni_ea_13}; (7) \citet{Motte_ea_98}; (8) \citet{Giannini_ea_12}\\
\smallskip
\end{table*}

\section{Discussion}
\label{Sec_Discussion}
\subsection{Time constraints on the model's applicability}
\label{Subsec_Time_constraints}
Basic presupposition of our model is a power-law $\rho$-PDF with constant slope which is independent on the spatial scale. In view of the definition of scales (formula \ref{def_abstr-scale}), this translates into the requirement in which the $\rho$-PDF is not subject to noticeable and/or stochastic changes in the considered time frame.  This leads to some constraints on evolutionary time or phase of the system.  Well developed, clearly distinguishable from lognormal wings PLTs of PDFs in self-gravitating clouds are to be expected at evolutionary times $\gtrsim 0.2 \tau_{\rm ff}$ \citep{Klessen_00, Girichidis_ea_14, Kritsuk_Norman_Wagner_11, Collins_ea_12}. Recent analysis of $\rho$-PDF PLTs in simulated self-gravitating clumps shows that the slope $|q|$ decreases smoothly within periods of $0.2 -2.5 \tau_{\rm ff}$, depending on the Jeans content, the initial velocity field and the type of turbulence driving, and suffers fast variations as it approaches the limiting value $\sim 1.5$ \citep{Veltchev_ea_19}. Taking also into account the zone of agreement in the CMF-parameter space between observations and our model (see Fig. \ref{fig_CMF_parameters}), we claim that the latter is best applicable at early phases of cloud collapse. Those are characterised by slopes $-2\gtrsim q \gtrsim -4$ and a temporary delay of the collapse due to stabilising agents like magnetic field and thermal pressure \citep[see discussion in][]{Girichidis_ea_14}. This physical picture is conceptually consistent with the result of our model: formation of massive, low-dense but super-Jeans cores which are subject to further fragmentation leading to protostellar objects. 

\subsection{Relation to other CMF models}
\label{Subsec_Comparison_with_other_models}
% \toadd{On similar models linking the cloud properties to the cores. }
Essentially, the presented model aims to reproduce the general CMF which is to be expected from turbulent or gravoturbulent fragmentation of dense clouds into condensations of different shapes and densities, modelled through statistical cores. In that sense, the model is similar to the approach of \citet{Padoan_Nordlund_02} who also derive the CMF slope through weighting over the scales in the fractal cloud (cf. their Sect. 5 with Sect. \ref{Sec_The CMF} in this paper). In their purely self-similar consideration, the obtained CMF slope is also $-1$, with no dependence on the fractal dimension of the cloud, i.e. on the local physical conditions. The main difference to the model of \citet{Padoan_Nordlund_02} is that we assume a purely power-law PDF (characteristic for dense protocluster clumps) in contrast to the lognormal PDF in their treatment. 

The model setting of \citet{Padoan_Nordlund_02} has been extended and upgraded to follow further the dynamical evolution of prestellar cores formed through initial cloud fragmentation. For instance, \citet{Dib_ea_07} investigate how the process of coa	lescence of cores in the inner part of MCs affect the time-evolution of the CMF and its transition to the stellar IMF. \citet{Dib_ea_10} take into account the internal structure of cores in terms of density profile and mass-density relationship whereas the core radius depends on the core mass and position in the cloud. These authors introduce a time-dependent accretion onto the cores which also depends on their location and leads to significant modification of the CMF in the course of cloud evolution. Alternatively, \citet{Dib_ea_13} make use of the same set of initial conditions in the cloud and study the evolution of cores without accretion but implementing the kinetic energy input by stellar winds from massive newly formed stars. Our model is comparable to the abovementioned works only in terms of the initial cloud fragmentation but without preference to any particular physical mechanism. The only assumption on the further core evolution is that unbound cores eventually dissolve while their bound peers contract and are being replenished. In regard to the mass function of unstable cores, the presented model follows the approach of \citet{CKB_07} who show that the initial (shallow) CMF should be corrected to account for the different free-fall times of cores and thus its slope would become similar to that of the Salpeter value of the stellar IMF or steeper.

All models referred above aim to connect the CMF with the stellar IMF although they do not discuss the origin of the CMF itself. The goal of this paper is not to reproduce the high-density slope of the IMF from the CMF. The treatment of the latter issue would require implementation of more physical processes, such as disk fragmentation or core merging, taking into account as well their variation in the star-forming environments. Rather -- sticking to the {\it statistical} approach of DVK17 -- we extend their study on possible links between general structure of MCs and characteristics of their fragments (cores) which eventually give birth to stars.

\subsection{Comparison with observational CMFs}
From an observational point of view, the proposed CMF model is applicable to MCs with large PLTs of their column-density distributions. The sole constraint stemming from the model is that the mean column densities of the extracted cores (regardless of the extraction method) are within the PLT range. Then, considering only the PLT, the column-density PDF translates -- under the model assumption of spherical symmetry -- to a power-law density PDF (see DVK17).

The range $-2\ge q \ge -4$  (leading to CMF slopes $-1.5\le \Gamma \le -1.17$) of good consistency of our model with observational CMFs (Fig. \ref{fig_CMF_parameters}) corresponds to values of the mass scaling index $\gamma$  between $1.5$ and $2.25$ (formula \ref{eq_M_ell-rho_ell}). Such range of $\gamma$ is entirely consistent with the general structure of Galactic molecular clouds  as studied in different tracers: molecular-line emissions, dust continuum and dust extinction. For instance, \citet{LAL_10} found  $\gamma\lesssim 2$, for a sample of clouds with large variation of masses, i.e. defined by different choice of column-density threshold. \citet{Kauffmann_ea_10} derived mass-size relationship with $\gamma=1.7$  for cloud fragments of sizes $1-4$ pc in several nearby star-forming regions.   The size and mass ranges in those works fit also well with the ones,  calculated from our model for a typical density contrast $\rho_0/\rho_{\rm c}=10^3$ and various values of the cloud's Jeans content: $0.6-$ 2 pc and $10^2 - 10^3~M_\odot$, respectively.	

The theoretical position of the characteristic mass\footnote{Or, "turnover mass" in case the CMF is modelled by a smooth function.} is a complex issue which is widely discussed in the literature. Considerations of Jeans collapse in turbulent medium show that $M_{\rm ch}$ depends only on global cloud parameters like sound speed and sonic scale, even when the assumption of isothermality is not valid \citep[see][and references therein]{Hopkins_12, Guszejnov_Hopkins_15}. On the other hand, \citet{Schmidt_ea_10} derive from simulations of supersonic isothermal turbulence CMFs of unstable cores with characteristic mass depending on the Jeans length in the numerical box and on (resolution) effects stemming from the applied clump-finding algorithm. The variation of $M_{\rm ch}$ from our model (determined by the quantity \ref{eq_Characteristic_mass}) is qualitatively comparable to their result. If the Jeans content of the cloud $M_{\rm c}/M_{\rm J,\,c}$ is comparable or exceeds within an order or of magnitude the number of extracted cores $N_{\rm c}$, the characteristic mass is consistent with observational works.

\section{Summary}
\label{Sec_Summary}
We propose a statistical model of the mass function of prestellar cores (CMF), generated at a given point of evolution of self-gravitating isothermal clouds. The latter are represented through abstract spherical objects characterised by single size, density profile, density contrast and parameters of the cloud core. The probability distribution function of mass density ($\rho$-PDF) in the cloud is assumed to be purely power law, with slope $q$.  The statistical prestellar cores are homogeneous spheres which populate cloud shells as determined by the corresponding log-density ranges.  

Basic assumptions of the model are power-law relations between core density, mass and size and self-similarity typical for fractal clouds. The main parameters are the total number of cores $N_{\rm c}$ (alternatively, the index of cloud fragmentation $1/\kappa^3$) and the cloud mass in Jeans masses $M_{\rm c}/M_{\rm J,\,c}$ (Jeans content of the cloud).

Our results are as follows:
\begin{enumerate}
\item  The CMF in general is a power law of slope $\Gamma=-1$. The found slope is to be expected if one considers the cores as hierarchical objects in a fractal cloud. It is in general agreement with a number of studies of the total clump population in star-forming regions.
\item Regarding the conditions for core collapse, the model yields three scenarios, conditioned by the PDF slope $q$.
\begin{itemize}
\item       $q=-3/2$ (well developed PL tail at advanced evolutionary stages):  all cores are either stable or unstable, depending on $N_{\rm c}$ and the Jeans content of the cloud. 
\item $-1>q>-3/2$: all cores {\it above} some threshold density collapse.
\item $q<-3/2$: all cores {\it below} some threshold density collapse. These are less dense, but massive objects, which are possibly subject to further fragmentation.
\end{itemize}
\item  The derived time-weighted CMF of gravitationally unstable cores is a power law of slope  $\Gamma=-1 + x/2$ where $x=1/(1+q)$ and $q\le-3/2$. This gives a good agreement with high-mass parts of observational CMFs for PDF slopes $-2 \ge q \ge -4$ which characterize earlier phases of cloud's collapse. The CMF of the total population in these cases is a combination of two power laws as the characteristic mass separates the regimes of non-collapsing and collapsing (high-mass) cores. 
\end{enumerate}

{\it Acknowledgement:}  
We are grateful to our anonymous referee for the critical and careful reading of the manuscript and for the valuable suggestions. S.D. acknowledges support by the Bulgarian National Science Fund under Grant N 12/11 (20.12.2017). T.V. acknowledges support by the DFG under grant KL 1358/20-1 and additional funding from the Ministry of Education and Science of the Republic of Bulgaria, National RI Roadmap Project DO1-277/16.12.2019, as well from the Scientific Research Fund of the University of Sofia, Grant \#80-10-68/19.04.2018. P.G. acknowledges funding from the European Research Council under ERC-CoG grant CRAGSMAN-646955. R.S.K. thanks funding from the DFG in the Collaborative Research Center (SFB 881) "The Milky Way System" (subprojects B1, B2, and B8) and in the Priority Program SPP 1573 "Physics of the Interstellar Medium" (grant numbers KL 1358/18.1, KL 1358/19.2).

\appendix
% \subsection{Statistical properties of the statistical objects}
\section{Statistical properties of cores}
\label{Appendix: Statistical properties}

\subsection{Distributions of masses, sizes and volumes}
\label{Appendix: Statistical distributions}

Those distributions are derived by:
\begin{eqnarray}
\label{eq_mass_distribution}
\begin{aligned}
p(s_{\rm m})ds_{\rm m} &= A_{\rm m}\exp(qxs_{\rm m})ds_{\rm m}&~ \\
 &= A_{\rm m} \left(\frac{m}{m_{\rm n}}\right)^{qx} d\ln\left(\frac{m}{m_{\rm n}}\right)~,&A_{\rm m}=xA_{\rm s}~,
\end{aligned}
\end{eqnarray}
\begin{eqnarray}
\label{eq_size_distribution}
\begin{aligned}
p(s_{\rm l})ds_{\rm l} &=A_{\rm l}\exp\left(q\frac{3x}{1-x}s_{\rm l}\right)ds_{\rm l}&~ \\
&= A_{\rm l} \left(\frac{l}{l_{\rm n}}\right)^{q\frac{3x}{1-x}} d\ln\left(\frac{l}{l_{\rm n}}\right)~,&A_{\rm l}=\frac{3x}{1-x}A_{\rm s}~,
\end{aligned}
\end{eqnarray}
\begin{eqnarray}
\label{eq_volume_distribution}
\begin{aligned}
p(s_{\rm v})ds_{\rm v} &=A_{\rm v}\exp\left(q\frac{x}{1-x}s_{\rm v}\right)ds_{\rm v}&~ \\
&= A_{\rm v} \left(\frac{v}{v_{\rm n}}\right)^{\frac{qx}{1-x}} d\ln\left(\frac{v}{v_{\rm n}}\right),&A_{\rm v}=\frac{x}{1-x}A_{\rm s}~.
\end{aligned}
\end{eqnarray}

Note that Eqs. (\ref{eq_mass_distribution})-(\ref{eq_volume_distribution}) above rest on the assumption that the power-law $\rho$-PDF is preserved at all scales compared to the core sizes. Indeed, this follows implicitly from the two assumptions (Eqs. (\ref{Equation: Density-mass-size relation})-(\ref{Equation: Core mass-size relation})), combined with the relation between the slope $q$ of the $\rho$-PDF and the structure parameter $x$ (Eq. (\ref{Equation: Structure parameter-slope relation})).

\subsection{Averaged quantities}
\label{Appendix: Averaged quantities}
The averaged core mass, density, volume and size over the entire cloud are obtained by use of the formulae obtained in Section \ref{Subsec_Model_Postulates}. According to the method of calculation, two types of averaging are distinguished: arithmetic and geometric (logarithmic). The obtained relations between the averaged quantities (Eqs. (\ref{arithm-averaged_relation-v}), (\ref{arithm-averaged_relation-l}) and (\ref{ln-averaged_relation})) could serve as tools to probe the applicability of the model for selected observed or simulated medium.

\subsubsection{Arithmetic averaged quantities}
\label{Subsec_Arithmetic_averaged}

The arithmetic average of core density is:
\begin{eqnarray}
\label{eq_rho_arithm-averaged}
\begin{aligned}
\overline{\left(\frac{\rho}{\rho_{\rm n}}\right)}_{\rm ar} &= A_{\rm s} \int\limits_{\rho_{\rm c}}^{\rho_0} \left(\frac{\rho}{\rho_{\rm n}}\right) \left(\frac{\rho}{\rho_{\rm n}}\right)^{q} d\ln\left(\frac{\rho}{\rho_{\rm n}}\right) = ... \\
& = \left(\frac{q}{q+1}\right) \left(\frac{\rho_{\rm c}}{\rho_{\rm n}}\right) \bigg[\frac{(\rho_0/\rho_{\rm c})^{1+q}-1}{(\rho_0/\rho_{\rm c})^{q}-1}\bigg]\simeq 1
\end{aligned}
\end{eqnarray}
This is a trivial result because $\rho_{\rm n}=\overline{\rho}_{\rm ar}$ and $\overline{\rho}_{\rm ar}\equiv \rho_{\rm n} \overline{(\rho/\rho_{\rm n})}_{\rm ar}$.

The arithmetic average of core mass is:
\begin{eqnarray}
\label{eq_m_arithm-averaged}
\begin{aligned}
\overline{\left(\frac{m}{m_{\rm n}}\right)}_{\rm ar} &= A_{\rm m} \int\limits_{\rho_{\rm c}}^{\rho_0} \left(\frac{m}{m_{\rm n}}\right) \left(\frac{m}{m_{\rm n}}\right)^{qx} d\ln\left(\frac{m}{m_{\rm n}}\right) = \\
& = \left(\frac{qx}{qx+1}\right) \left(\frac{m_{\rm c}}{m_{\rm n}}\right) \left[\frac{(\rho_0/\rho_{\rm c})^{\frac{1}{x}+q}-1}{(\rho_0/\rho_{\rm c})^{q}-1}\right]\simeq \\
&\simeq \left(\frac{q}{2q+1}\right)\!\!\left(\frac{m_{\rm c}}{m_{\rm n}}\right) =\!\left(\frac{q}{2q+1}\right)\!\! \left(\frac{1+q}{q}\right)^{1+q},
\end{aligned}
\end{eqnarray}
where at the last step we make use of $m_{\rm c}/m_{\rm n}=(\rho_{\rm c}/\rho_{\rm n})^{1+q}=\left((1+q)/q\right)^{1+q}$ (cf. Eqs. (\ref{Equation: Mass-density relation}), (\ref{Equation: Normalization coefficient of size}) and (\ref{Equation: Structure parameter-slope relation})).

The arithmetic average of core volume is:
\begin{eqnarray}
\label{eq_v_arithm-averaged}
\begin{aligned}
\overline{\left(\frac{v}{v_{\rm n}}\right)}_{\rm ar} &= A_{\rm v} \int\limits_{\rho_{\rm c}}^{\rho_0} \left(\frac{v}{v_{\rm n}}\right) \left(\frac{v}{v_{\rm n}}\right)^{q\frac{x}{1-x}} d\ln\left(\frac{v}{v_{\rm n}}\right) = ...\\
&= \left(\frac{q\frac{x}{1-x}}{1+q\frac{x}{1-x}}\right) \left(\frac{v_{\rm c}}{v_{\rm n}}\right) \bigg[\frac{(\rho_0/\rho_{\rm c})^{\frac{1-x}{x}+q}-1}{(\rho_0/\rho_{\rm c})^{q}-1}\bigg] \\
& \simeq \frac{1}{2} \left(\frac{v_{\rm c}}{v_{\rm n}}\right) = \frac{1}{2} \left(\frac{1+q}{q}\right)^{q}~,
\end{aligned}
\end{eqnarray}
where at the last step we make use of $v_{\rm c}/v_{\rm n}=(\rho_{\rm c}/\rho_{\rm n})^{q}=((1+q)/q)^{q}$.

And the arithmetic average of the size reads:
\begin{eqnarray}
\label{eq_l_arithm-averaged}
\begin{aligned}
\overline{\left(\frac{l}{l_{\rm n}}\right)}_{\rm ar} &= A_{\rm l} \int\limits_{\rho_{\rm c}}^{\rho_0} \left(\frac{l}{l_{\rm n}}\right) \left(\frac{l}{l_{\rm n}}\right)^{q\frac{3x}{1-x}} d\ln\left(\frac{l}{l_{\rm n}}\right) = ... \\
& = \left(\frac{q\frac{3x}{1-x}}{1+q\frac{3x}{1-x}}\right) \left(\frac{l_{\rm c}}{l_{\rm n}}\right) \bigg[\frac{(\rho_0/\rho_{\rm c})^{\frac{1-x}{3x}+q}-1}{(\rho_0/\rho_{\rm c})^{q}-1}\bigg] \\
& \simeq \frac{3}{4} \left(\frac{l_{\rm c}}{l_{\rm n}}\right) = \frac{3}{4} \left(\frac{1+q}{q}\right)^{q/3}~,
\end{aligned}
\end{eqnarray}
where at the last step we make use of $l_{\rm c}/l_{\rm n}=(\rho_{\rm c}/\rho_{\rm n})^{q/3}=((1+q)/q)^{q/3}$.

Eventually we define the arithmetic average quantities as follows:  $\overline{\rho}_{\rm ar}\equiv \rho_{\rm n} \overline{(\rho/\rho_{\rm n})}_{\rm ar}$, $\overline{m}_{\rm ar}\equiv m_{\rm n} \overline{(m/m_{\rm n})}_{\rm ar}$, $\overline{v}_{\rm ar}\equiv v_{\rm n} \overline{(v/v_{\rm n})}_{\rm ar}$ and $\overline{l}_{\rm ar}\equiv l_{\rm n} \overline{(l/l_{\rm n})}_{\rm ar}$. From the Eqs. (\ref{eq_rho_arithm-averaged}), (\ref{eq_m_arithm-averaged}), (\ref{eq_v_arithm-averaged}) and (\ref{eq_l_arithm-averaged}) are derived the relations:
\begin{eqnarray}
\label{arithm-averaged_relation-v}
\frac{\overline{m}_{\rm ar}}{\overline{\rho}_{\rm ar} \overline{v}_{\rm ar}}= 2 \left(\frac{1+q}{1+2q}\right)~,
\end{eqnarray}
\begin{eqnarray}
\label{arithm-averaged_relation-l}
\frac{\overline{m}_{\rm ar}}{\overline{\rho}_{\rm ar} \frac{4\pi}{3}\overline{l}_{\rm ar}^3}= \left(\frac{4}{3}\right)^3 \left(\frac{1+q}{1+2q}\right)~.
\end{eqnarray}
Note the difference between last two formulas. In contrast to it, the analogous relations derived for the corresponding geometric averaged quantities turned out to be equal (see the end of the next Subsection).

\subsubsection{Geometric (logarithmic) averaged quantities}
\label{Subsec_Geometric_averaged}

The geometric average of core density is defined:
\begin{eqnarray} 
\overline{\left(\frac{\rho}{\rho_{\rm n}}\right)}_{\rm ln}\equiv \exp(\overline{\ln(\rho/\rho_{\rm n})})~~, \nonumber \end{eqnarray}
where
\[ \overline{\ln\left(\frac{\rho}{\rho_{\rm n}}\right)}= A_{\rm s}  \int\limits_{\rho_{\rm c}}^{\rho_0} \ln\left(\frac{\rho}{\rho_{\rm n}}\right) \left(\frac{\rho}{\rho_{\rm n}}\right)^{q} d\ln\left(\frac{\rho}{\rho_{\rm n}}\right)~~. \]
Then one obtains after some calculations:
\begin{eqnarray}
\label{eq_rho_ln-averaged}
\begin{aligned}
\overline{\left(\frac{\rho}{\rho_{\rm n}}\right)}_{\rm ln}&=...= \\
\exp\left(-\frac{1}{q}\right) & \left(\frac{\rho_0}{\rho_{\rm n}}\right)^{\frac{(\rho_0/\rho_{\rm n})^{q}}{(\rho_0/\rho_{\rm n})^{q}-(\rho_{\rm c}/\rho_{\rm n})^{q}}}\!\! \left(\frac{\rho_{\rm c}}{\rho_{\rm n}}\right)^{-\frac{(\rho_{\rm c}/\rho_{\rm n})^{q}}{(\rho_0/\rho_{\rm n})^{q}-(\rho_{\rm c}/\rho_{\rm n})^{q}}} \\
& \simeq \exp\left(-\frac{1}{q}\right) \left(\frac{\rho_{\rm c}}{\rho_{\rm n}}\right)= \exp\left(-\frac{1}{q}\right) \left(\frac{1+q}{q}\right)
\end{aligned}
\end{eqnarray}

The geometric average of core mass is defined in analogous way:
\begin{eqnarray}
\overline{\left(\frac{m}{m_{\rm n}}\right)}_{\rm ln}\equiv \exp(\overline{\ln(m/m_{\rm n})})~~. \nonumber
\end{eqnarray}
The use of the relationship $\ln(m/m_{\rm n})= (1/x)\ln(\rho/\rho_{\rm n})$ and, hence,  $\overline{\ln(m/m_{\rm n})}= (1/x)\overline{\ln(\rho/\rho_{\rm n})}$ simplifies the calculation of the averaged core mass:
\begin{eqnarray}
\label{eq_m_ln-averaged}
\overline{\left(\frac{m}{m_{\rm n}}\right)}_{\rm ln}= \bigg[\overline{\left(\frac{\rho}{\rho_{\rm n}}\right)}_{\rm ln}\bigg]^{1/x}\simeq \bigg[\exp\left(-\frac{1}{q}\right) \left(\frac{1+q}{q}\right)\bigg]^{1/x}~.
\end{eqnarray}

The geometric average core volume and size are defined and calculated analogously:
\begin{eqnarray}
\label{eq_v_ln-averaged}
\begin{aligned}
\overline{\left(\frac{v}{v_{\rm n}}\right)}_{\rm ln} & \equiv \exp(\overline{\ln(v/v_{\rm n})})= \left[\overline{\left(\frac{\rho}{\rho_{\rm n}}\right)}_{\rm ln}\right]^{\frac{1-x}{x}} \\
& \simeq \bigg[\exp\left(-\frac{1}{q}\right) \left(\frac{1+q}{q}\right)\bigg]^{\frac{1-x}{x}}~~, 
\end{aligned}
\end{eqnarray}
\begin{eqnarray}
\label{eq_l_ln-averaged}
\begin{aligned}
\overline{\left(\frac{l}{l_{\rm n}}\right)}_{\rm ln}& \equiv \exp(\overline{\ln(l/l_{\rm n})})= \left[\overline{\left(\frac{\rho}{\rho_{\rm n}}\right)}_{\rm ln}\right]^{\frac{1-x}{3x}} \\
& \simeq \bigg[\exp\left(-\frac{1}{q}\right) \left(\frac{1+q}{q}\right)\bigg]^{\frac{1-x}{3x}}~~. 
\end{aligned}
\end{eqnarray}

Eventually we define the geometric averaged quantities as follows: $\overline{\rho}_{\rm ln}\equiv \rho_{\rm n} \overline{(\rho/\rho_{\rm n})}_{\rm ln}$, $\overline{m}_{\rm ln}\equiv m_{\rm n} \overline{(m/m_{\rm n})}_{\rm ln}$, $\overline{v}_{\rm ln}\equiv v_{\rm n} \overline{(v/v_{\rm n})}_{\rm ln}$ and $\overline{l}_{\rm ln}\equiv  l_{\rm n} \overline{(l/l_{\rm n})}_{\rm ln}$. From Eqs. (\ref{eq_rho_ln-averaged}), (\ref{eq_m_ln-averaged}), (\ref{eq_v_ln-averaged}) and (\ref{eq_l_ln-averaged}) we derive a relationship between geometric averaged density, mass, volume and size of the cores:
\begin{eqnarray}
\label{ln-averaged_relation}
\frac{\overline{m}_{\rm ln}}{\overline{\rho}_{\rm ln} \overline{v}_{\rm ln}}=\frac{\overline{m}_{\rm ln}}{\overline{\rho}_{\rm ln} \frac{4\pi}{3}\overline{l}_{\rm ln}^3}= 1~.
\end{eqnarray}

\label{lastpage}

\end{document}